\documentstyle[aps,preprint]{revtex}
\begin{document}
 
\title{Imprints of log-periodic self-similarity in the stock market}
 
\author{S. Dro\.zd\.z$^{1,2}$, F. Ruf$^{3}$, J. Speth$^{1}$,
and M. W\'ojcik$^{1,2}$}
\address{
$^{1}$Institut f\"ur Kernphysik, Forschungszentrum J\"ulich,
D-52425 J\"ulich, Germany \\
$^{2}$ Institute of Nuclear Physics, PL-31-342 Krak\'ow, Poland\\
$^{3}$ WestLB International S.A., 32-34 bd Grande-Duchesse Charlotte,
L-2014 Luxembourg}
\date{\today}
\maketitle

\begin{abstract}
Detailed analysis of the log-periodic structures as precursors of the financial
crashes is presented. 
The study is mainly based on the German Stock Index (DAX) 
variation over the 1998 period which includes both, 
a spectacular boom and a large decline, in magnitude only comparable
to the so-called Black Monday of October 1987. 
The present example provides further arguments in favour of a discrete
scale-invariance governing the dynamics of the stock market.
A related clear log-periodic structure prior to the crash and
consistent with its onset extends over the period of a few months. 
Furthermore, on smaller time-scales the data seems to indicate 
the appearance of analogous log-periodic oscillations as precursors
of the smaller, intermediate decreases. Even the frequencies of such 
oscillations are similar on various levels of resolution.
The related value $\lambda \approx 2$ of preferred scaling
ratios is amazingly consistent with those found for a wide variety of other 
complex systems. Similar analysis of the major American indices between 
September 1998 and February 1999 also provides some evidence supporting 
this concept but, at the same time, illustrates a possible splitting of the
dynamics that a large market may experience.

\end{abstract}
 
 
\bigskip
 
\newpage

The fact that a healthy and normally functioning financial market may 
reveal certain properties common to complex systems 
is fascinating and, in fact, seems natural.
Especially interesting in this context is the recently suggested 
analogy of the financial crashes to critical points 
in statistical mechanics~\cite{Sorn1,Feig1,Vand,Feig2,Gluz,Joha1}.
Criticality implies a scale invariance which in mathematical terms, 
for a properly defined function $F(x)$ characterizing the system,
means that for small $x$
\begin{equation}
F(\lambda x) = \gamma F(x).
\label{eq:F}
\end{equation}
A positive constant $\gamma$ in this equation describes how the properties
of the system change when it is rescaled by the factor $\lambda$.  
The simplest solution to this equation reads:
\begin{equation}
F_{0}(x) = x^{\alpha},
\label{eq:pow}
\end{equation}
where $\alpha = \log(\gamma)/\log(\lambda)$. This is a standard power-law that
is characteristic of continuous scale-invariance and $\alpha$ is the 
corresponding critical exponent. 

More interesting is the general solution~\cite{Naue} to Eq.~(\ref{eq:F}):
\begin{equation}
F(x) = F_{0}(x) P({\log F_{0}(x) / \log(\gamma)}),
\label{eq:logper}
\end{equation}
where $P$ denotes a periodic function of period one.
In this way the dominating scaling (\ref{eq:pow}) acquires a correction
which is periodic in $\log(x)$.
This solution accounts for a possible discrete scale-invariance~\cite{Sorn2}
and can be interpreted~\cite{Sorn3,Newm} in terms of a complex critical 
exponent
$\alpha = \alpha_R + i \alpha_I$, since
$\Re \{x^{\alpha}\} = x^{\alpha_R} \cos(\alpha_I \log(x))$, which corresponds
to the first term in a Fourier expansion of (\ref{eq:logper}).
Thus, if $x$ represents a distance to the critical point, the resulting
spacings between consecutive minima $x_n$ (maxima) of the log-periodic 
oscillations seen in the linear scale follow a geometric contraction
according to the relation:
\begin{equation}
{x_{n+1} - x_n \over x_{n+2} - x_{n+1}} = \lambda.
\label{eq:gp}
\end{equation}
Then, the critical point coincides with the accumulation of such oscillations.

Existence of the log-periodic modulations correcting the structureless 
pure power-law behaviour has been identified in many different
systems~\cite{Sorn2}. 
Examples include diffusion-limited-aggregation clusters~\cite{Sorn4},
crack growth~\cite{Ball}, earthquakes~\cite{Sorn3,Newm} and, 
as already mentioned, the financial market 
where $x$ is to be interpreted as the time to crash.
Especially in the last two cases this is an extremely interesting 
feature because it potentially offers a tool for predictions.    
Of course, the real financial market is exposed to many external factors
which may distort its internal hierarchical structure on the organizational
as well as on the dynamical level. Therefore, the searches for the long term,
of the order of few years, precursors of crashes have to be taken 
with some reserve, as already pointed out in Ref.~\cite{Lalo}. 
A somewhat related example is shown in Fig.~1 which displays 
the S$\&$P 500 versus DAX charts between 1991 and February 1999. 
While the global characteristics of the two charts are largely compatible
there exist several significant differences on shorter time-scales.
It is the purpose of the present paper to explore more in detail the 
emerging short-time behaviour of the stock market indices. 

On the more general ground, the current attitude in developing
the related theory is logically not fully consistent. 
First of all, no methodology is provided as how to 
incorporate a pattern of log-periodic oscillations preceding a particular 
crash into an even more serious crash which potentially may occur in a year 
or two later. 
Secondly, even though there is some indication that the largest crashes
are outliers by belonging to a different population~\cite{Joha2},
there exists no precise definition of what is to be 
qualified as a crash, especially in the context of its analogy to critical 
points of statistical mechanics. Just a bare statement that the crash 
corresponds to discontinuity in the derivative of an appropriate market index
is not sufficiently accurate to decide what amount of decline 
is needed to signal a real crash and what is to be 
considered only a 'correction'. In fact, a closer inspection
of various maket indices on different time-scales suggests that 
it is justifiable to consider them as nowhere differentiable. An emerging 
scenario of the market evolution, in a natural way resolving this kind of
difficulties, would then correspond to 
a permanent competition between booms and crashes of various sizes; a picture
somewhat analogous to the self-organized critical state~\cite{Bak}
and consistent with a causal information cascade from large scales to small 
scales as demonstrated through the analysis 
of correlation functions~\cite{Arne}.
In this connection the required existence of many critical points 
within the renormalization group theory may result from a more general 
nonlinear renormalization flow map, 
i.e. by replacing $\lambda x$ by $\phi(x)$~\cite{Derr,Sale}. 
In fact, such a mechanism may even remain compatible with the log-periodic
scaling properties of Eq.~(\ref{eq:logper}) on various scales.
For this to apply the accumulation
points of the log-periodic oscillations on smaller scales need themselves   
to be distributed as the log-periodic sequence.
  
Identification of a clean hierarchy of the above suggested structures 
on the real market is not expected to be an easy task
because of a possible contamination by various external factors or by some 
internal market nonuniformities. 
However, on the longer time-scales many such factors may  
cancel out to a large extent .
Within the shorter time-intervals, on the other hand, the influence of such 
factors can significantly be reduced by an appropriate selection of the  
location of such intervals.
In this later sense one finds the most preferential 
conditions in the recent DAX behaviour as no obvious external events 
that may have influenced its evolution can be indicated.  
Here, as illustrated in Fig.~2, within the period of only 9 month 
preceding July 1998 the index went up from about 3700 to almost 6200 
and then quickly declined to below 4000. This draw down is however somewhat
slower than some of the previous crashes analysed in similar context 
but for this reason it even better resembles a real physical second order
phase transition. During the spectacular boom period
the three most pronounced deep minima, indicated by the upward long solid-line 
arrows, can immediately be located. Denoting the resulting times as 
$t_n$, $t_{n+1}$, $t_{n+2}$ and making the correspondence with
Eq.~(\ref{eq:logper}) by setting $x_i =t_c -t_i$, where $t_c$ 
is the crash time, already such three points can be used to determine $t_c$:  
\begin{equation}
t_c = {t^2_{n+1} - t_{n+2} t_n \over 2 t_{n+1} - t_n - t_{n+2}}.
\label{eq:tc}
\end{equation}
The result is indicated by a similar downward arrow and reasonably 
well agrees with the actual time of crash. 
The corresponding preferred scaling ratio between
$t_{n+1} - t_n$ and $t_{n+2} - t_{n+1}$ (Eq.~(\ref{eq:gp})), governing
the log-periodic oscillations, gives $\lambda=2.17$, which is consistent 
with the previous cases analysed in the literature
not only in connection with the market evolution~\cite{Joha3}
but for a wide variety of other systems as well~\cite{Sorn2}
and, thus, may indicate a universal character of the mechanism responsible
for discrete scale invariance in complex systems.

As a further element of the present analysis it can be quite clearly seen 
from Fig.~2 
that there is essentially no qualitative difference between the nature 
of the major crash and those index declines that mark its preceding
log-periodically distributed minima. Indeed, they also seem to be preceded
by their own log-periodic oscillations within appropriately shorter
time-intervals. The two such sub-sequences are indicated by the long-dashed
and short-dashed arrows, respectively, and the corresponding $t_c$ calculated
using Eq.~(\ref{eq:tc}) by downward arrows of the same type. 
In both cases the so-estimated
$t_c$'s also reasonably well coincide with times of the decline.
Interestingly, the scaling ratios $\lambda$ for these log-periodic structures
equal 2.06 and 2.07, respectively, and thus turn out consistent  
with the above value of 2.17. Moreover, even on the deeper level
of resolution the two sequences of identifiable oscillations indicated 
by the dotted-line arrows in Fig.~2 develop analogous structures resulting  
in $\lambda=2.26$ (earlier case) and $\lambda=2.1$ (later case), 
which is again consistent with all its previous values. 
This means that such a whole fractal hierarchy of log-periodic
structures may still remain grasped by one function of the
type~(\ref{eq:logper}). 
In this scenario the largest crash of Fig.~2 may appear as 
just one component of log-periodic oscillations extending into the future
and announcing an even larger future crash.
Large crashes can thus be assigned no particular role in this sense. 
They are preceded by the log-periodic oscillations of about the same 
frequency as the small ones. What still can make them outliers~\cite{Joha2}
are parallel secondary effects like an overall increase of the market volume
which may lead to an additional amplification of their amplitude.    
    
Violant reverse in a market tendency may reveal log-periodic-like structures
even during the intra-day trading. One such example is illustrated in Fig.~3
which shows the minutely DAX variation between 11:50 and 15:30 on January 8,
1999. It is precisely during this period (still seen in Fig.~1) that DAX 
reached its few month maximum after recovery from the previously 
discussed crash. Taking the average of the ratios between the consecutive
five neighbouring time-intervals determined by the     
six deepest minima indicated by the upward arrows results here in 
$\lambda \approx 1.7$. The corresponding $t_c$ (downward arrow) 
again quite precisely indicates the onset of the decline.      

Directly before the major crash on July 20, 1998 the trading dynamics
somewhat slows down (as can be seen from Fig.~2) and no such structures on the
level of minutely variation can be identified in this case, however. 
In fact, a fast increase of the market index just 
before its subsequent decline seems to offer the most 
favourable conditions for the log-periodic oscillations to show up on 
the time-scales of a few month or shorter. This may simply reflect the fact 
that a faster internal market dynamics generates such oscillations of larger
amplitude which thus gives them a better chance to dominate a possible external
corruption. Consistently, another market (Hong-Kong Stock Exchange) 
whose Hang Seng index went up recently by almost $40\%$ during about 
a four month period in 1997 also provides quite a convincing example 
of the short term log-periodic oscillations. The four most pronounced minima 
at 97.24, 97.43, 97.52 and 97.56 trace a geometric progression with a common 
$\lambda$ of about 2.15 and this progression converges to $t_c \approx 97.6$, 
thus exactly indicating the begining of a dramatic crash.             
The relevant Hang Seng chart can be seen in Fig.~2 of ref.~\cite{Joha1}.
      
Instead of listing further examples where the log-periodic oscillations
accompany a local fast increase of the market index we find it more instructive
to study more in detail the recent development on the American market.
It provides further evidence in favour of this concept but at the same time
illustrates certain possible related subtleties. Fig.~4 shows the S$\&$P500 
behaviour starting mid September 1998 versus the two closely related 
and most frequently quoted indices: 
the Dow Jones Industrial Average (DJIA) which is entirely comprised 
by S$\&$P500 and the Nasdaq-100 whose about $60\%$ of the volume 
overlaps with S$\&$P500. The latter two indices (DJIA and Nasdaq-100) 
are totally disconnected in terms of the company content.
In order to make the relative speed of the changes 
directly visible all these indices are normalized such that they
are equal to unity at the same date (here on February 1, 1999).     
Clearly, it is Nasdaq which within the short period between October 8, 1998 
(its lowest value in the period considered) and February 1, 1999 develops
a very spectacular rise by almost doubling its magnitude. In this case the
three most pronounced consecutive minima ($\lambda=2.25$) also quite precisely
point to the onset of the following 11$\%$ correction. Parallel increase of the
DJIA is much slower, the pattern of oscillations much more difficult 
to uniquely interpret and, consistently, no correction occurs.
At the same time the S$\&$P500 largely behaves like an average of the two.
Even though it displays similar three minima as Nasdaq, the early February 
correction is only rudimentary. In a sense we are thus facing an example of 
a very interesting temporary spontaneous decoupling of a large market, 
as here represented by S$\&$P500, into submarkets some of which  
may evolve for a certain period of time according to their own 
log-periodic pattern of oscillations which are masked in the global index.  
The smaller, more uniform markets are less likely to experience such effects
of decoupling and are thus expected to constitute better candidates 
to manifest the short-time universal structures. The opposite may apply
to the global index since from the longer time-scales 
perspective such effects should be less significant. The examples analysed 
here as well as those studied in the literature are in fact consistent
with this interpretation.  
If applies, such an interpretation provides another physically 
appealing picture: short-time log-periodicity is more localized in the
'market space' while the longer-time-scales probe its more global aspects.

Tabulating the critical exponents $\alpha_R$ for 
the stock market in the context of our present study of criticality on various
time-scales doesn't seem equally useful as their values significantly 
depend on the time-window inspected. For instance, in the case of the 
DAX index we find values ranging between 0.2 for the few years-long
time-intervals up to almost 1 for the shorter time-intervals corresponding
to the identified log-periodic substructures.
One may argue that the logarithm of the stock market index 
constitutes a more appropriate quantity for determining the critical exponents.
However, also on the level of the logarithm the exponents vary between 
about 0.3 up to 1 in analogous time-intervals as above.

In conclusion, the present analysis provides further arguments for
the existence of the log-periodic oscillations constituting a significant
component in the time-evolution of the fluctuating part 
of the stock market indices. 
Even more, imprints are found for the whole hierarchy of such log-periodically
oscillating structures on various time-scales and this hierarchy carries
signatures of self-similarity. 
An emerging scenario of the market evolution characterized by nowhere
differentiable permanent competition between booms and crashes of various size 
is then much more logically acceptable and consistent.     
Of course, in general, it would be naive to expect that on the real
market any index fluctuation can uniquely be classified as a member 
of a certain log-periodically distributed sequence.  
Some of such fluctuations may be caused by external factors which are likely
to be completely random relative to the market intrinsic evolutionary
synchrony. 
It is this complex intrinsic interaction of the market constituents which     
may lead to such universal features as the ones discussed above and it is
extremely interesting to see that such features (a consistent sequence of 
the log-periodically distributed oscillations) can quite easily be 
identified with help of some physics guidance.      
The above result makes also clear that the stock market log-periodicity 
reveals much richer structure than just lowest order Fourier expansion
of Eq.~(\ref{eq:logper})~\cite{Sorn1,Feig1,Vand}, and therefore, 
at the present stage the 'arrow dropping' procedure used here offers much
more flexibility in catching the essential structures 
and seems thus more appropriate, especially on shorter time-scales. 
Finally, in this context we wish to draw attention to the 
Weierstrass-Mandelbrot fractal function     
\footnote{The Weierstrass-Mandelbrot fractal function is defined as
$W(t) = \sum_{n = -\infty}^{\infty} 
{(1 - e^{i\gamma^nt})e^{i\phi_n} / \gamma^{\eta n} }$, 
where $0 < \eta < 1$, $\gamma < 1$ and $\phi_n$ is an arbitrary phase.
It is easy to show by relabeling the series index that for an appropriate
choise of the set of phases $\{\phi_n\}$,
$W(\gamma t) = \gamma^{\eta} W(t)$.}
which is continuous everywhere,
but is nowhere differentiable~\cite{Berr,Sorn2} and can be made to obey the
renormalization group equation. The relevance of this function 
for log-periodicity has already been pointed out in connection with 
earthquakes~\cite{Boro}. 
It is likely that a variant of this function also provides an appropriate 
representation for the stock market criticality.

\bigskip 

We thank R. Felber and Dr. M. Feldhoff for very useful discussions 
on the related matters. 
We also wish to thank Dr. A. Johansen and Prof. D. Sornette for their useful
comments on the earlier version (cond-mat/9901025) of this paper and for 
bringing to our attention some of the references that we were unaware of.

\newpage
\begin{center}
{\bf FIGURE CAPTIONS}
\end{center}
{\bf Fig.~1.} The Deutsche Aktienindex--DAX (upper chart) 
versus S\&P~500 (lower chart) in the period 1991-1999.
Logarithms of both indices are shown for a better comparison within the same
scale. \\
{\bf Fig.~2.} The daily evolution of the Deutsche Aktienindex from
October 1997 to October 1998. Upward arrows indicate minima of 
the log-periodic oscillations used to determine the corresponding critical   
times denoted by the downward arrows. Different types of arrows 
(three upward and one downward) correspond to different sequences of 
log-periodic oscillations identified on various time-scales. \\ 
{\bf Fig.~3.} The minutely DAX variation between 11:50 and 15:30 
on January 8, 1999. Upward arrows indicate minima used to determine 
the corresponding critical time. \\
{\bf Fig.~4} The daily variation of S$\&$P500 (upper panel), 
Nasdaq-100 (middle panel) and Dow Jones Industrial Average (lower panel) 
from September 1998 till March 1999. All these indices are normalized such
they equal unity on February 1, 1999. \\
 
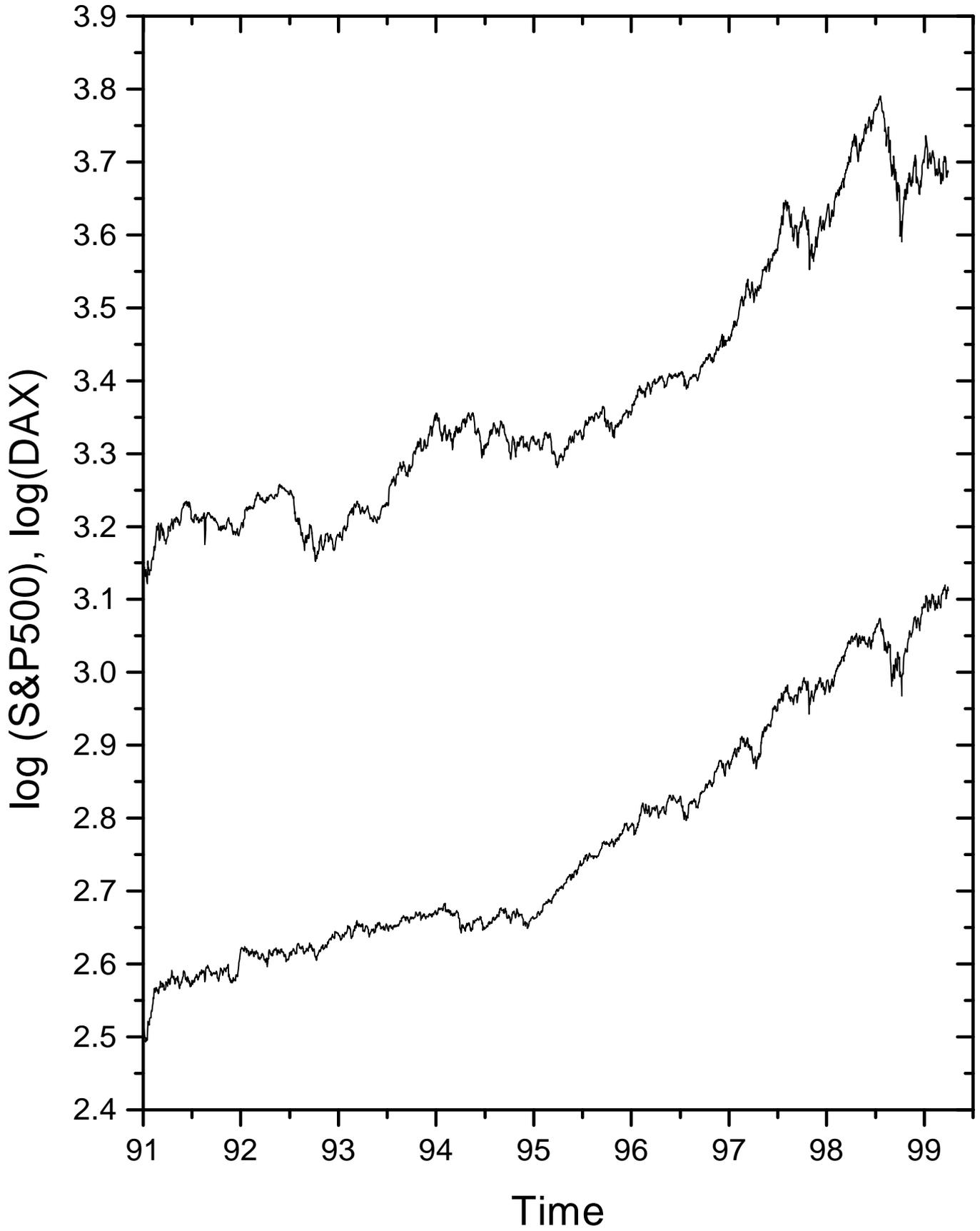
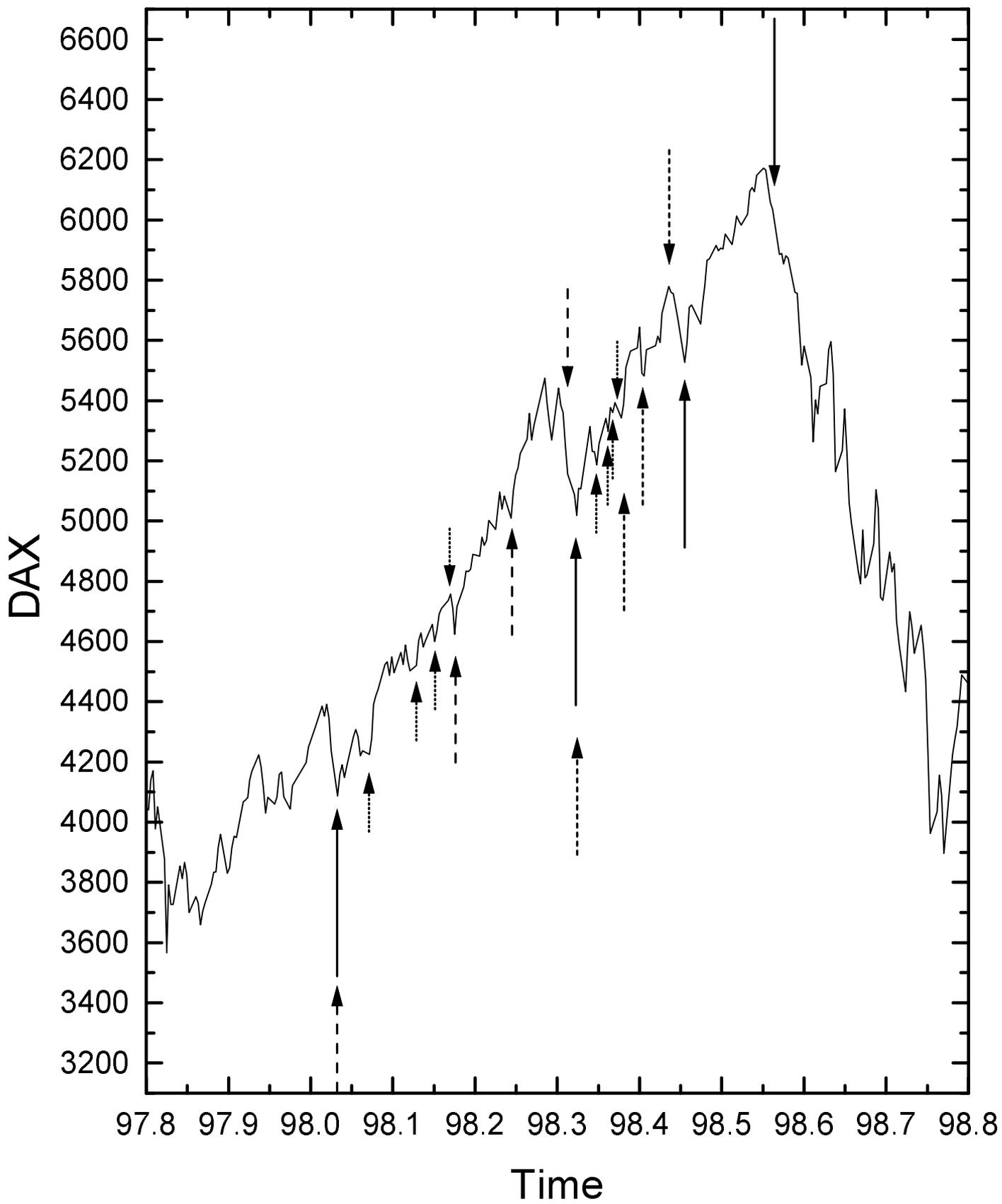
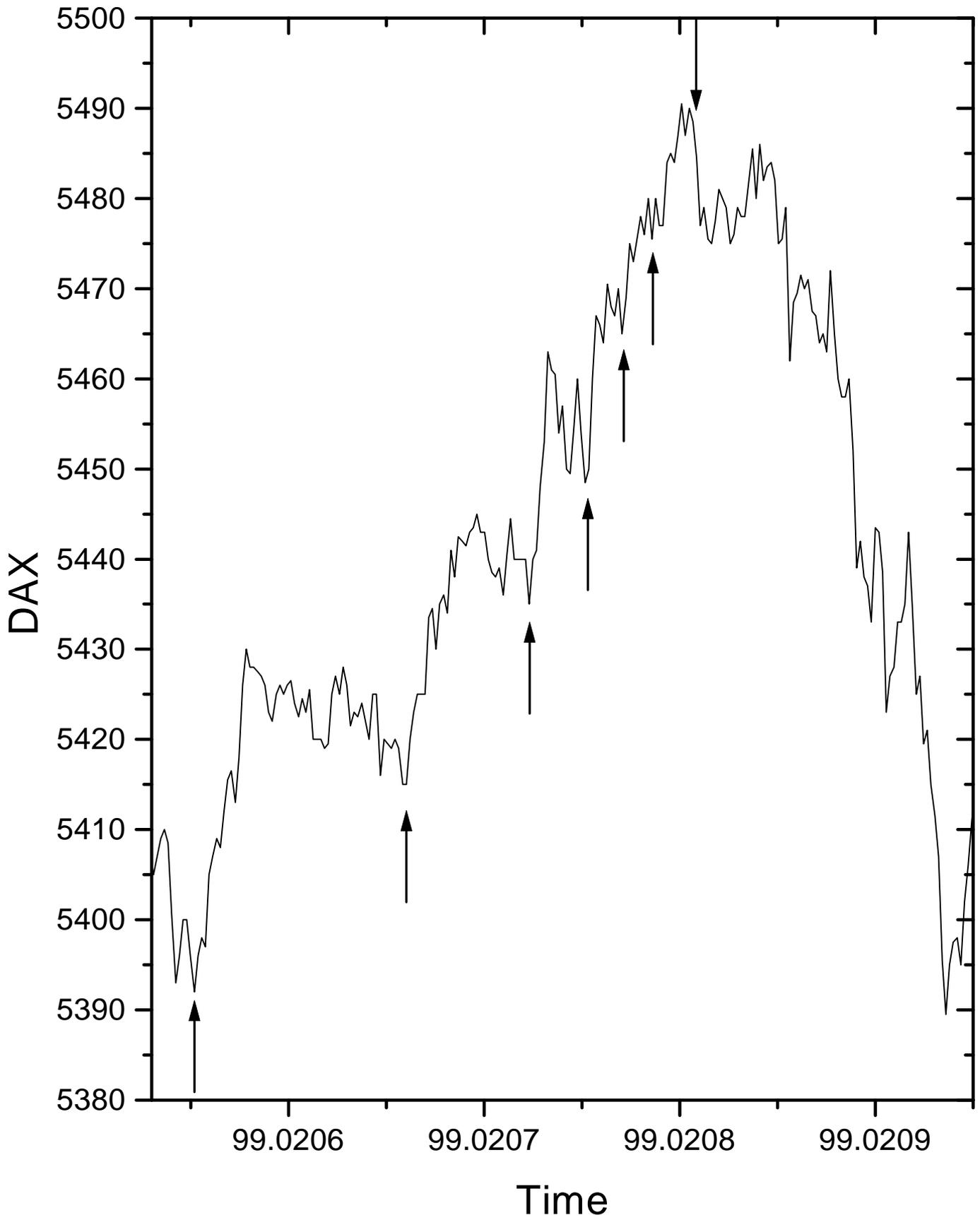
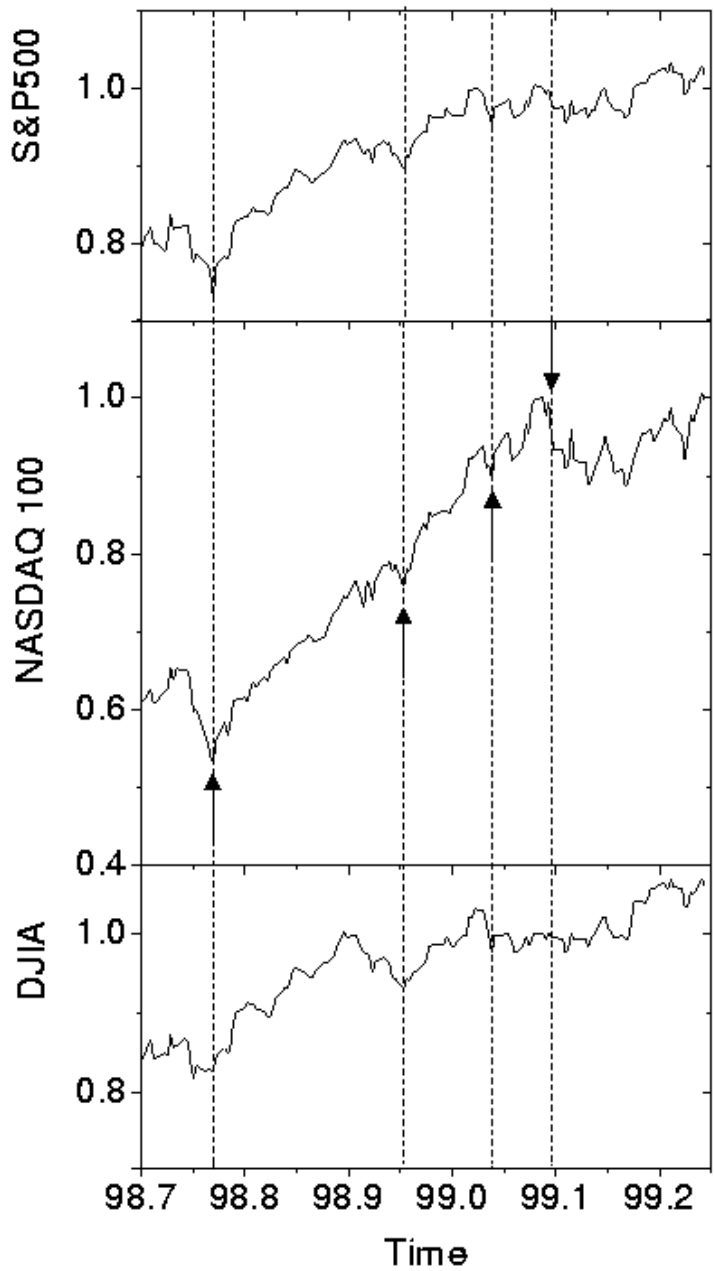
\end{document}